\documentclass[12pt,english,svgnames]{article}
\usepackage[utf8]{inputenc}
\usepackage[english]{babel}
\usepackage{float}
\usepackage{booktabs}
\usepackage{amsmath,amssymb,amsthm,amsfonts}
\usepackage{enumerate}
\usepackage{tikz,graphicx,xcolor,caption}
\usepackage[authoryear,round]{natbib}
\usepackage[hyphens]{url}
\usepackage{hyperref}
\usepackage[a4paper,body={16cm,23.7cm}]{geometry}

\usetikzlibrary{positioning}

\hypersetup{colorlinks=true, linkcolor=cyan, citecolor=cyan, urlcolor=black, breaklinks=true}

\newtheorem{theorem}{Theorem}

\newtheorem{proposition}{Proposition}

\begin{document}

\title{\textbf{Proportional redistribution}\thanks{The authors gratefully acknowledge grants PID2023-147391NB-I00 and PID2023-146364NB-I00, funded by MCIU/AEI/10.13039/501100011033 and FSE+.}
}
\author{\textbf{Ricardo Mart\'{\i}nez}\thanks{ Universidad de Granada,
Spain. email: ricardomartinez@ugr.es} \and \textbf{Juan D. Moreno-Ternero}\thanks{Universidad Pablo de Olavide, Spain. email: jdmoreno@upo.es}}
\maketitle

\begin{abstract}
The ethic of proportional redistribution is a compromise between the extremely compensatory ethic of full redistribution and the needs-blind ethic of laissez-faire. In a basic model of redistribution problems with needs, we characterize proportional redistribution with a combination of axioms that formalize minimal requirements of accountability, functionality, and impartiality. 
Consequently, we provide a new ethical rationalization for the ancient Aristotelian maxim of proportionality. 
\end{abstract}

\bigskip

\medskip {}


\newpage

\section{Introduction}
Redistribution is arguably the most important task of all governments. As \cite{tullock2013economics} puts it, ``\textit{it seems quite probable that the original motive for forming governments was the desire for special types of redistribution}". But there exist many aspects that influence the specific redistributive policy chosen in a society. A large literature studies some of them, which range from lingustic diversity to pork, religion, xenophobia or the composition of the political class \citep[e.g.,][]{Lee2006, Desmet2009, Huber2013, Corvalan2018, Gallice2019}. In particular, it is well known that different beliefs about the fairness of social competition and what determines income inequality might play a major role \citep[e.g.,][]{Alesina2005, Durante2014, almaas2020cutthroat}. Societies in which individual effort is thought to determine income will typically defend low redistribution, whereas those giving a more prominent role to luck will typically defend high redistribution. 
Those views are taken to the extreme in non-democratic societies \citep[e.g.,][]{boix2003democracy}.\footnote{Redistribution may also play an important role in transition to democracy and the eventual emergence and persistence of inefficient states based on patronage politics \citep[]{Acemoglu2011}} In a right-wing authoritarian regime, the poor are excluded from the decision-making process, which drives towards \textit{laissez-faire}. In a communist regime, the wealth of the rich is confiscated, allegedly to implement \textit{full redistribution}. 

In this paper, we are going to make a case in favor of \textit{proportional redistribution}, according to agents' needs. The notion can be connected to the ancient Aristotelian maxim of proportionality \citep[e.g.,][]{crisp2014aristotle}: ``\textit{In a just distribution, unequal individuals should get unequal shares, in proportion to their unequal worth or merit}". This principle has nowadays overwhelming support for a wide variety of settings in Western societies \citep[e.g.,][]{CohenEliya2011, Bansak2017, Cappelen2019}. In our setting, proportional redistribution obviously constitutes an intermediate position between the extreme options of laissez-faire and full redistribution, accounting for the heterogeneity of individuals. In that sense, it is also reminiscent of the classical notion of \textit{equal sacrifice} that was extremely influential in the early days of the literature on public finance \citep[e.g.,][]{young1990progressive}. 
Finally, as we shall show here, it has strong normative foundations too.  

We consider a stylized model of a society, in which its members are only described by two numbers. One of them is \textit{income} (or performance), to be interpreted as the individual contribution to the wealth of the society, which could be negative. The other is \textit{need}, which is a non-negative number to be interpreted as an objective requirement the individual has, to be possibly covered from the wealth of the society. This means that an individual's income may or may not be above the individual's need and, also, that the wealth of the society might not be enough to fully cover the overall need society has. A \textit{rule} is a mapping associating to each society so described (i.e., to each pair of income and need profiles) an allocation of income, indicating how much each individual gets from the wealth of the society after redistribution. 

In the framework described above, we consider four axioms formalizing properties of rules with normative appeal (either from an ethical, or an operational viewpoint). The first axiom is indeed a package of three \textit{core} principles: homogeneity (scale invariance), equal treatment of equals (in terms of incomes and needs), and continuity. The other axioms are \textit{no advantageous transfer} (no group of agents can benefit from reallocating incomes and needs among their members), \textit{stability} (after a rule redistributes income, it requests no further redistribution for the resulting income profile and \textit{dummy} (an agent with null income and needs gets nothing). 

Our main result states that the combination of those (apparently innocuous) axioms characterizes exactly two rules in our setting: laissez-faire and proportional redistribution. This is a remarkable result which provides strong normative foundations for proportional redistribution (laissez-faire would be easily dismissed, for instance, adding any axiom formalizing that redistribution is not fully independent on needs). An operational (in the form of continuity, homogeneity and stability) and impartial (in the form of equal treatment of equals) redistribution rule that imposes minimal requirements of accountability (in the form of dummy) and prevention of coalitional manipulations (in the form of no advantageous transfer) that is not laissez-faire, has to be proportional.

Our work can be seen as part of the literature on \textit{axiomatic analysis of resource allocation} \citep[e.g.,][]{Moulin1997}, whose origins are the (welfarist) theories of bargaining \citep[]{Nash1950} and coalitional form games \citep[]{Shapley1953}, as well as the (resourcist) theory of fair division \citep[]{foley1966resource}. The latter is closer to our setting, as it only involves a vector of commodities (the physical resources to be divided) and a profile of preference orderings. 
The unidimensional version of fair division typically excludes preferences from the model, and identifies the resource as money.\footnote{This is implicitly assuming that preferences are the same for all agents, that the allocation only pays attention to the monetary amounts agents receive, or that they have linear utility over how much they receive \citep[e.g.,][]{Thomson2023}.} 
This is the case of the problem of adjudicating conflicting claims introduced by \cite{ONeill1982}, or its dual taxation problem \citep[e.g.,][]{Young1988}.\footnote{\cite{moreno2006impartiality, MorenoTernero2012} also study a unidimensional resource allocation problem, but in which agents are endowed with capability functions to transform the resource into outcome.} 
If the amount to allocate is zero, and agents are allowed to obtain negative amounts, then we have the so-called redistribution problems \citep[e.g.,][]{ju2007non, Casajus15}. Our extension 
to account for needs is akin to extensions of standard models such as income inequality \citep[e.g.,][]{Almaas2011, Hufe2022}, cost sharing \citep[e.g.,][]{de2023accounting}, or claims problems \citep[e.g.,][]{Hougaard2012}. 

The rest of the paper is organized as follows. In Section \ref{model}, we introduce the model and the basic definitions (including our axioms) for our analysis. In Section~\ref{main results}, we provide our characterization theorems, which give normative support for proportional redistribution. In Section~\ref{further}, we provide further insights, exploring notions such as duality and income additivity, which give rise to extra characterizations. We conclude in Section~\ref{conclusion}. For a smooth passage, we defer the proofs of all results to an Appendix.
\section{The model}\label{model}
Let $N=\{1,\dots, n\}$ be a set of individuals, or \textbf{agents}. For each $i\in N$, let $y_{i}\in \mathbb{R}$ and $z_{i}\in \mathbb{R}_+$ be $i$'s \textbf{income} and \textbf{need}, respectively.\footnote{Note that incomes may be negative.} We denote by $y\equiv (y_i)_{i\in N}$ and $z\equiv (z_i)_{i\in N}$ the corresponding profiles of incomes and needs. The aggregate income is $Y \equiv \sum_{i \in N} y_{i}$, while the aggregate need is $Z \equiv \sum_{i \in N} z_{i}$. We assume that $Z>0$. A redistribution problem with needs (in short, a \textbf{problem}) is a pair $(y,z) \in \mathbb{R}^N \times \mathbb{R}_+^N$. Let $\mathcal{D}^{N}$ be the domain of all problems. We shall also consider a variable-population generalization of the model. Then, there is a set of potential agents, which are indexed by the natural numbers $\mathbb{N}$.  Let $\mathcal{N}$ be the set of finite subsets of $\mathbb{N}$, with generic element $N$. Let $\mathcal{D} \equiv \bigcup _{N\in \mathcal{N}}\mathcal{D}^{N}$ be the domain of all problems. For any $N' \subset N$, $y_{N'}$ and $z_{N'}$ denote the restriction of the vectors $y$ and $z$ to $N'$. Similarly, $Y_{N'} = \sum_{i \in N'} y_i$ and $Z_{N'} = \sum_{i \in N'} z_i$ are the aggregate income and need of agents in $N'$. Besides, for any $i \in N$, $y_{-i} = y_{N\backslash i}$, $z_{-i} = z_{N\backslash i}$, $Y_{-i} = Y_{N\backslash i}$, $Z_{-i} = Z_{N\backslash i}$.

Given a problem $(y,z) \in \mathcal{D}^{N}$, an \textbf{allocation} is a vector of real numbers $x \equiv (x_i)_{i \in N} \in \mathbb{R}^N$ such that $\sum_{i \in N} x_i = Y$. Let $X(y,z)$ denote the set of all allocations for the problem $(y,z)$. An allocation rule, or simply a \textbf{rule}, is a mapping $R:\mathcal{D} \longrightarrow \bigcup_{N \in \mathcal{N}} \mathbb{R}^N$ that selects, for each problem $(y,z) \in \mathcal{D}^{N}$, a unique allocation $R(y,z) \in X(y,z)$.

Throughout the rest of the paper we assume that each rule satisfies the following pack of basic (\textbf{core}) conditions. The first one is the usual condition of scale invariance, stating that if we multiply incomes and needs by a constant factor, so is the solution to the problem. The second one is the standard impartiality requirement stating that agents with the same income and needs should receive the same amount. The third one is the natural condition that small changes in the data of the problem should only produce small changes in the outcome of the problem. Formally,

\begin{itemize}
    \item[(i)] \textbf{Homogeneity}. For each $(y,z) \in \mathcal{D}^N$ and $\rho \in \mathbb{R}_{++}$, $R(\rho y , \rho z)= \rho R(y,z)$.
    \item[(ii)] \textbf{Equal treatment of equals}. For each $(y,z) \in \mathcal{D}^{N}$ and each pair $i,j \in N$, if $y_i=y_j$ and $z_i=z_j$, then $R_i(y,z)=R_j(y,z)$.
    \item[(iii)] \textbf{Continuity}. For each sequence $\{(y^\nu,z^\nu)\}$ of problems in $\mathcal{D}^{N}$ and each $(y,z) \in \mathcal{D}^{N}$, if $\{(y^\nu,z^\nu)\}$ converges to $(y,z)$, then the sequence $\{R(y^\nu,z^\nu)\}$ converges to $R(y,z)$.
\end{itemize}

We simply write that a rule \textit{satisfies core} when it satisfies homogeneity, equal treatment of equals and continuity. 

Two focal and polar rules (that obviously satisfy core) are the following. First, the standard laissez-faire, which amounts to no redistribution whatsoever. Formally, 

\textbf{laissez-faire}. For each $(y,z) \in \mathcal{D}^{N}$ and each $i \in N$, 
$$
R^{L}_i(y,z)=y_{i}.
$$

In contrast, we can also formalize the rule that implies (full) confiscation of income, to allocate it equally among agents. Formally, 

\textbf{Full redistribution}. For each $(y,z) \in \mathcal{D}^{N}$ and each $i \in N$,
$$
R^{F}_i(y,z)=\frac{Y}{n}.
$$

A natural (and less radical) alternative appeals to the long-standing Aristotelian principle of proportionality, according to which agents obtain a portion of the overall income that is proportional to their individual needs. Formally, 

\textbf{Proportional redistribution}. For each $(y,z) \in \mathcal{D}^{N}$ and each $i \in N$, 
$$
R^{P}_i(y,z)= \frac{z_i}{Z} \, Y.
$$

Apart from the \textit{core axioms}, we shall consider three main axioms in our work (some of which might allow us to distinguish among the previous and other rules). First, a standard notion that has a long tradition in axiomatic work to prevent strategic manipulations in resource allocation \citep[e.g.,][]{Moulin1985, Moulin1987, ju2007non}. It states that no group of agents can benefit from reallocating incomes and needs among their members. Formally, 

\textbf{No advantageous transfer}. For each pair $(y,z),(y',z') \in \mathcal{D}^{N}$ and each $N' \subset N$, if $Y_{N'}=Y'_{N'}$, $Z_{N'}=Z'_{N'}$, $y_{N \backslash N'}=y'_{N \backslash N'}$ and $z_{N \backslash N'}=z'_{N \backslash N'}$, then
$$
\sum_{i \in N'} R_i(y,z) = \sum_{i \in N'} R_i(y',z').
$$

The second axiom is a natural requirement of stability in this setting \citep[e.g.,][]{Chambers2021, Martinez2022}. It says that after a rule redistributes income, then it proposes no further redistribution for the resulting income profile. In other words, the rule is idempotent. Formally,

\textbf{Stability}. For each $(y,z) \in \mathcal{D}^N$,
$$
R(R(y,z),z)=R(y,z).
$$

The third axiom is also a classical requirement in axiomatic work, that can be traced back to \cite{Shapley1953}. It states that an agent with null income and needs gets nothing in the allocation process.

\textbf{Dummy}. For each $(y,z) \in \mathcal{D}^N$ and each $i \in N$, if $y_i=z_i=0$ then $R_i(y,z)=0$.

\section{The main results}\label{main results}
Our first result states that the combination of all the axioms introduced above characterizes a pair of rules; namely, laissez-faire and the proportional rule. 
\begin{theorem}\label{LF+PROP}
A rule satisfies core, no advantageous transfer, stability and dummy if and only if it is either laissez-faire or the proportional rule.
\end{theorem}
Theorem \ref{LF+PROP} is remarkable as axioms that are seemingly innocuous independently, end up having a strong bite together. Not only that, as they end up characterizing a pair of rules that are completely different. On the one hand, the rule that fully ignores the structure of the model and any possible form of redistribution, thus yielding the initial distribution as the solution. On the other hand, the rule that imposes a redistribution that is proportional to individual needs. That is why we interpret this result as strong normative support for proportional redistribution. 

As we shall see in the next results, dismissing just one of the axioms in the previous combination drastically widens the scope of rules that become admissible. Common to all these rules will be a weighing factor that depends on the \textit{social cost of needs}, to be formalized as the ratio of overall income and need, i.e., $\frac{Z}{Y}$.  

\begin{theorem}\label{NAT+ST}
A rule $R$ satisfies core, no advantageous transfer, and stability if and only if it is laissez-faire, or there exists a continuous function $B: \mathbb{R} \longrightarrow \mathbb{R}$, such that, for each $(y,z) \in \mathcal{D}^N$ and each $i \in N$,
$$
R_i(y,z)= \frac{Y}{n} + \left[z_i-\frac{Z}{n} \right] B \left( \frac{Y}{Z} \right).
$$
\end{theorem}
Theorem \ref{NAT+ST} shows that dismissing dummy widens the scope to allow for rules that generalize proportional redistribution (apart from laissez-faire that, obviously, remains a valid option). This generalized proportionality admits a very elegant functional form that combines full redistribution with a weighted version of \textit{needs differential}. That is, each individual tentatively receives first an equal share of the aggregate income, and this amount is adjusted up or down, depending on whether his (individual) need is below or above the average need. The adjustment is weighted by a function of the social cost of needs ($\frac{Y}{Z}$). Note that if this function is the identity then we precisely have the proportional rule. If it is not, we have generalizations of it. 

\begin{theorem}\label{NAT+D}
A rule $R$ satisfies core, no advantageous transfer, and dummy if and only if there exists a continuous function $A: \mathbb{R} \longrightarrow \mathbb{R}$, such that, for each $(y,z) \in \mathcal{D}^N$ and each $i \in N$,
$$
R_i(y,z)= A \left( \frac{Y}{Z} \right) y_i + \left[ 1- A \left( \frac{Y}{Z} \right)\right] \frac{Y}{Z}z_i.
$$

\end{theorem}
Theorem \ref{NAT+D} shows that dismissing stability instead widens the scope to allow for a convex combination of proportional redistribution and laissez-faire, with the weighing factor being a function of the social cost of needs ($\frac{Y}{Z}$). Note that if this function is null then we precisely have the proportional rule, whereas if it is constant and equal to $1$ we precisely have laissez-faire. Any other function gives rise to compromises between the two rules. 

\begin{theorem}\label{thm_AB}
A rule satisfies core and no advantageous transfer if and only if 
there exist a pair of continuous functions $A,B: \mathbb{R} \longrightarrow \mathbb{R}$, such that for each $(y,z) \in \mathcal{D}^N$ and each $i \in N$,
$$R^{AB}_i(y,z)= \frac{Y}{n} +  \left[y_i-\frac{Y}{n} \right] A \left( \frac{Y}{Z} \right) + \left[z_i-\frac{Z}{n} \right] B \left( \frac{Y}{Z} \right).$$
\end{theorem}

Finally, if we get rid of both axioms and only keep no advantageous transfer (together with core) we end up characterizing a large family of rules that compromises among laissez-faire, \textit{needs differential} and \textit{incomes differential}. That is, each individual tentatively receives first an equal share of the aggregate income, and this amount is adjusted up or down depending on whether his (individual) need and income are below or above the average need and income. The adjustment is weighted by two arbitrary functions of the social cost of needs ($\frac{Y}{Z}$). Note that if both functions are null, then we precisely have full redistribution. If one function is null and the other is constant and equal to 1, then we precisely have laissez-faire, or need-adjusted full redistribution. 
Finally, if both functions are constant we have a linear combination of full redistribution, laissez-faire or need-adjusted full redistribution that we have characterized elsewhere \citep[e.g.,][]{Martinez2024}.




\section{Further insights}\label{further}
It is well known that individuals weigh gains and losses differently \citep[e.g.,][]{Tversky1979}. This motivates a dual approach in our setting, in which agents rather focus on the allocation of the income gap, and its differences with respect to the needs vector.\footnote{Duality has been widely explore in related contexts \citep[e.g.,][]{Young1988, Thomson2008, Oishi2016, Atay2026}.} 
That is, for each rule we can define its dual as follows.

\textbf{Dual of rule $R$}. For each $(y,z) \in \mathcal{D}^{N}$, $R^d(y,z)=z-R(z-y,z)$.

It is straightforward to see that laissez-faire and proportional redistribution are self-dual rules. That is, the dual rule is the rule itself (i.e., $R^d \equiv R$). The dual rule of full redistribution is the following rule, which was implicitly mentioned above. 

\textbf{Need-adjusted full redistribution}. For each $(y,z) \in \mathcal{D}^{N}$ and each $i \in N$, 
$$
R^{N}_i(y,z)= z_i + \frac{Y-Z}{n}.
$$
Finally, the dual of each rule within the family described at Theorem \ref{thm_AB} is also a rule within the same family. More precisely, $({R^{AB}})^d\equiv R^{A^dB^d}$, where $A^d(t)=A(1-t)$ and $B^d(t)=1-A(1-t)-B(1-t)$.

Analogously, we say that two axioms are dual of each other if, whenever a rule satisfies one of them, its dual rule satisfies the other. All the axioms considered above (namely, core, no advantageous transfer, stability and dummy) are self-dual. The following axiom (which is satisfied by laissez-faire and proportional redistribution, among others) is not. 

\textbf{Income additivity}. For each pair $(y,z),(y',z') \in \mathcal{D}^{N}$,
$$
R(y+y',z)=R(y,z)+R(y',z)
$$
Its dual is the following.

\textbf{Dual of income additivity}. For each pair $(y,z),(y',z') \in \mathcal{D}^{N}$,
$$
z+R(y+y',z)=R(y,z)+R(z+y',z)
$$
The next result states that adding income additivity to core and no advantageous transfer we characterize the linear combination among laissez-faire, full redistribution and proportional redistribution. If, instead, we add its dual axiom, we characterize (as expected) the dual family of the previous one. That is, we characterize the linear combination among laissez-faire, need-adjusted full redistribution and proportional redistribution.  

\begin{proposition}\label{thm_LPF} The following statements hold:
\begin{itemize}
    \item A rule $R$ satisfies core, no advantageous transfer, and income additivity if and only if there exists $(\alpha_1,\alpha_2) \in \mathbb{R}^2$ such that, for each $(y,z) \in \mathcal{D}^N$ and each $i \in N$,
$$
R_i(y,z)=\alpha_1 y_i + \alpha_2 \frac{z_i}{Z} Y + (1-\alpha_1-\alpha_2) \frac{Y}{n}.
$$
\item A rule $R$ satisfies core, no advantageous transfer, and the dual of income additivity if and only if there exists $(\alpha_1,\alpha_2) \in \mathbb{R}^2$ such that, for each $(y,z) \in \mathcal{D}^N$ and each $i \in N$,
$$
R_i(y,z)=\alpha_1 y_i + \alpha_2 \frac{z_i}{Z} Y + (1-\alpha_1-\alpha_2) \left[ z_i+ \frac{Y-Z}{n} \right].
$$
\end{itemize}
\end{proposition}
If we add stability to the previous statements, we derive interesting selections within those families. 
That is, apart from laissez-faire, we characterize the families that arise as convex combinations between proportional redistribution and either full redistribution or its dual. 
\begin{proposition}\label{thm_PF} The following statements hold:
\begin{itemize}
    \item A rule satisfies core, no advantageous transfer, stability and income additivity if and only if it is either laissez-faire or a convex combination between full redistribution and proportional redistribution. 
\item A rule satisfies core, no advantageous transfer, stability and the dual of income additivity if and only if it is either laissez-faire or a convex combination between need-adjusted full redistribution and proportional redistribution.
\end{itemize}
\end{proposition}
If we add dummy instead of stability, we obtain a unique selection, made of the convex combinations between laissez-faire and proportional redistribution.  
\begin{proposition}\label{LP-cvx} 
A rule satisfies core, no advantageous transfer, dummy and income additivity (or its dual) if and only if it is a convex combination between laissez-faire and proportional redistribution. 
\end{proposition}

Finally, if we add income additivity (or its dual) to the whole package of axioms (core, no advantageous transfer, stability, and dummy) nothing changes, as the two rules characterized with that package (namely, laissez-faire and proportional redistribution) already satisfy income additivity (and its dual). 

\section{Conclusion}\label{conclusion}
In this paper, we have provided normative foundations for proportional redistribution. In a basic setting of redistribution problems with needs, we have shown that such a rule is essentially the only impartial and operational redistribution rule that imposes minimal requirements of accountability and prevention of coalitional manipulations. We have also obtained alternative characterizations (typically involving generalized versions of proportional redistribution) upon exploring various combinations of those axioms, the implications of a few extra axioms, as well as the notion of duality. This has allowed us to uncover the structure of the domain of redistribution problems with needs. 

The model we consider here is a generalization of the income redistribution model introduced by \cite{ju2007non} in which the problem is addressed without considering needs. 
They focus on the notion of \textit{reallocation-proofness} (a group of agents cannot manipulate the outcome of the rule upon reallocating the incomes within the group) and show that this axiom, together with \textit{no transfer paradox} (transferring some income before redistribution takes place does not increase income after redistribution), 
characterizes the family of income-tax schedules with a flat tax rate and personalized lump-sum transfers. That is, a sort of \textit{generalized proportional} rules, including as specific members the convex combinations between laissez-faire and full redistribution, characterized independently by \cite{Casajus15} and \cite{Martinez2022}.\footnote{\cite{Chambers2021} characterize \textit{threshold} rules for the two-agent case of this model, that guarantee partial redistribution for unequal incomes.}

\section*{Appendix}\label{Appendix}
We provide in this appendix all the proofs of the results stated above. We start with the most general result in which we only consider the axioms of \emph{core} and \emph{no advantageous transfer}, as it will pave the way for the remaining proofs. 

\subsection*{Proof of Theorem \ref{thm_AB}}
We prove first the straightforward implication. Let $A,B: \mathbb{R} \longrightarrow \mathbb{R}$ be two continuous functions. 
\begin{itemize}
    \item \emph{Core}. Let $(y,z) \in \mathcal{D}^N$ and $\rho \in \mathbb{R}_{++}$. For each $i \in N$,
    $$
    R^{AB}_i(\rho y,\rho z)= \frac{\rho Y}{n} +  \rho \left[y_i-\frac{Y}{n} \right] A \left( \frac{\rho Y}{\rho Z} \right) + \rho \left[z_i-\frac{Z}{n} \right] B \left( \frac{\rho Y}{\rho Z} \right) = \rho R^{AB}_i(y,z).
    $$
    And thus, $R^{AB}$ satisfies condition (i). Now, let $(y,z) \in \mathcal{D}^{N}$ and $i,j \in N$ be such that $y_i=y_j$ and $z_i=z_j$. Then, 
    $$
    R^{AB}_i(y,z) = \frac{Y}{n} +  \left[y_i-\frac{Y}{n} \right] A \left( \frac{Y}{Z} \right) + \left[z_i-\frac{Z}{n} \right] B \left( \frac{Y}{Z} \right) = R^{AB}_j(y,z).
    $$
    Then, $R^{AB}$ satisfies condition (ii). Finally, condition (iii) trivially follows from the fact that $A$ and $B$ are continuous functions. 
    \item \emph{No advantageous transfer}. Let $(y,z),(y',z') \in \mathcal{D}^{N}$ and $N' \subset N$ such that $Y_{N'}=Y'_{N'}$, $Z_{N'}=Z'_{N'}$, $y_{N \backslash N'}=y'_{N \backslash N'}$ and $z_{N \backslash N'}=z'_{N \backslash N'}$. Then,
    \begin{align*}
    \sum_{i \in N'} R^{AB}_i (y,z) &=  \sum_{i \in N'} \left[ \frac{Y}{n} +  \left[y_i-\frac{Y}{n} \right] A \left( \frac{Y}{Z} \right) + \left[z_i-\frac{Z}{n} \right] B \left( \frac{Y}{Z} \right) \right] \\
    &= \sum_{i \in N'} \left[ \frac{Y}{n} -\frac{Y}{n} A \left( \frac{Y}{Z} \right) -\frac{Z}{n} B \left( \frac{Y}{Z} \right) \right] - A \left( \frac{Y}{Z} \right) Y_{N'} - B \left( \frac{Y}{Z} \right) Z_{N'}  \\
    &= \sum_{i \in N'} \left[ \frac{Y'}{n} -\frac{Y'}{n} A \left( \frac{Y'}{Z'} \right) -\frac{Z'}{n} B \left( \frac{Y'}{Z'} \right) \right] - A \left( \frac{Y'}{Z'} \right) Y'_{N'} - B \left( \frac{Y'}{Z'} \right) Z'_{N'}  \\
    & = \sum_{i \in N'} \left[ \frac{Y'}{n} +  \left[y'_i-\frac{Y'}{n} \right] A \left( \frac{Y'}{Z'} \right) + \left[z'_i-\frac{Z'}{n} \right] B \left( \frac{Y'}{Z'} \right) \right] \\
    &= \sum_{i \in N'} R^{AB}_i (y',z').
    \end{align*}
\end{itemize}
We now focus on the converse implication. Let $R$ be a rule that satisfies \emph{core} and \emph{no advantageous transfer}. We divide the proof into several steps. 

\textbf{Step 1}. Let $i \in N$ and let $(y,z) \in \mathcal{D}$. If we take $N'=N \backslash i$, \emph{no advantageous transfer} requires that $\sum_{k \in N'} R_k(y,z)$ only depends on $Y_{N'}$ and $Z_{N'}$. As $R_i(y,z)=Y - \sum_{k \in N'} R_k(y,z)$, $Y_{N'}=Y-y_i$ and $Z_{N'}=Z-z_i$, we conclude that $R_i(y,z)$ only depends on the values of $y_i$, $z_i$, $Y$ and $Z$. Therefore, for each $i \in N$, let us define the function $\Delta_i$ as:
$$
\Delta_i(y_i,z_i,Y,Z)=R_i(y,z).
$$
\textbf{Step 2}. Now, we take $N'=\{i,j\}$ as a proper subset of any pair of agents in $N$. Let $(y,z),(y',z') \in \mathcal{D}^N$ be any two problems such that $y'_i=y_j$, $y'_j=y_i$, $z'_i=z_j$, $z'_j=z_i$, $y'_{N \backslash N'} = y_{N \backslash N'}$, and $z'_{N \backslash N'} = z_{N \backslash N'}$. By \emph{no advantageous transfer}, 
\begin{align*}
\Delta_i(y_i,z_i,Y,Z) + \Delta_j(y_j,z_j,Y,Z) &= \Delta_i(y'_i,z'_i,Y',Z') + \Delta_j(y'_j,z'_j,Y',Z') \\
&= \Delta_i(y_j,z_j,Y,Z) + \Delta_j(y_i,z_i,Y,Z)
\end{align*}
Equivalently,
$$
(\Delta_i - \Delta_j)(y_i,z_i,Y,Z) = (\Delta_i - \Delta_j)(y_j,z_j,Y,Z).
$$
On the one hand, the previous expression implies that $(\Delta_i - \Delta_j)$ is independent of the particular values of $(y_i,z_i)$. On the other hand, \emph{core} implies 
\begin{equation}\label{eq1}
R_i \left( \left( \frac{Y}{n}, \ldots, \frac{Y}{n} \right), \left( \frac{Z}{n}, \ldots, \frac{Z}{n} \right) \right) = R_j \left( \left( \frac{Y}{n}, \ldots, \frac{Y}{n} \right), \left( \frac{Z}{n}, \ldots, \frac{Z}{n} \right) \right) = \frac{Y}{n}
\end{equation}
Therefore, for any pair $(y_i,z_i)$,
\begin{align*}
(\Delta_i - \Delta_j)(y_i,z_i,Y,Z) &= (\Delta_i - \Delta_j) \left( \frac{Y}{n}, \frac{Z}{n} ,Y,Z \right) \\
&= R_i \left( \left( \frac{Y}{n}, \ldots, \frac{Y}{n} \right), \left( \frac{Z}{n}, \ldots, \frac{Z}{n} \right) \right) - R_j \left( \left( \frac{Y}{n}, \ldots, \frac{Y}{n} \right), \left( \frac{Z}{n}, \ldots, \frac{Z}{n} \right) \right) \\
&=0.
\end{align*}
In other words, as $\Delta_i=\Delta_j$ for each pair $(y_i,z_i)$, $\Delta_i$ is independent of agent $i$, and then we can simply suppress the subindex and write that, for each $i \in N$, 
$$
R_i(y,z) = \Delta (y_i,z_i,Y,Z).
$$
\textbf{Step 3}. Let $(y,z) \in \mathcal{D}^N$. \emph{No advantageous transfer} implies that 
\begin{multline*}
R_i\left( \left( y_i, y_j, y_{N \backslash \{i,j\}} \right), \left( z_i, z_j, z_{N \backslash \{i,j\}} \right) \right) + R_j\left( \left( y_i, y_j, y_{N \backslash \{i,j\}} \right), \left( z_i, z_j, z_{N \backslash \{i,j\}} \right) \right) = \\ 
R_i\left( \left( \frac{y_i+y_j}{2}, \frac{y_i+y_j}{2}, y_{N \backslash \{i,j\}} \right), \left( \frac{z_i+z_j}{2}, \frac{z_i+z_j}{2}, z_{N \backslash \{i,j\}} \right) \right) \\ + R_j\left( \left( \frac{y_i+y_j}{2}, \frac{y_i+y_j}{2}, y_{N \backslash \{i,j\}} \right), \left( \frac{z_i+z_j}{2}, \frac{z_i+z_j}{2}, z_{N \backslash \{i,j\}} \right) \right).
\end{multline*}
Besides, \emph{core} implies that 
\begin{multline*}
R_i\left( \left( \frac{y_i+y_j}{2}, \frac{y_i+y_j}{2}, y_{N \backslash \{i,j\}} \right), \left( \frac{z_i+z_j}{2}, \frac{z_i+z_j}{2}, z_{N \backslash \{i,j\}} \right) \right) = \\ 
R_j\left( \left( \frac{y_i+y_j}{2}, \frac{y_i+y_j}{2}, y_{N \backslash \{i,j\}} \right), \left( \frac{z_i+z_j}{2}, \frac{z_i+z_j}{2}, z_{N \backslash \{i,j\}} \right) \right).
\end{multline*}
Therefore, by definition of $\Delta$, it follows that
$$
\Delta \left( y_i,z_i,Y,Z \right) + \Delta \left( y_j,z_j,Y,Z \right)= 2 \Delta \left( \frac{y_i+y_j}{2},\frac{z_i+z_j}{2},Y,Z \right).
$$
Now, if we define the function $f: \mathbb{R} \times \mathbb{R}_+ \longrightarrow \mathbb{R}$ as $f(r,s) = \Delta \left(r,s,Y,Z \right)$, we have 
$$
\frac{f(r_1,s_1) + f(r_2,s_2)}{2}=  f\left( \frac{r_1+r_2}{2},\frac{s_1+s_2}{2} \right),
$$
which is 
Jensen's equation. As $R$ is continuous, it follows by \emph{core} that 
$\Delta$ is also continuous, and thus so is $f$. The solution to Jensen's equation is then $f(r_1,s_1)=ar_1 + bs_1 + c$ (e.g. Aczel (2006), pag. 218), which implies that 
$$
R_i(y,z)=\Delta (y_i,z_i,Y,Z) = a(Y,Z)y_i + b(Y,Z)z_i + c,
$$
where $a,b: \mathbb{R} \times \mathbb{R}_+ \longrightarrow \mathbb{R}$ are two continuous functions (as $R$ is), and $c \in \mathbb{R}$ is a constant.
By (\ref{eq1}), we can obtain the expression of the constant $c$, and thus,
\begin{equation}\label{eq2}
R_i(y,z)= \frac{Y}{n} +  \left[y_i-\frac{Y}{n} \right] a(Y,Z) + \left[z_i-\frac{Z}{n} \right] b(Y,Z).
\end{equation}
\textbf{Step 4}. Notice that, in the previous expression of $R_i(y,z)$, the functions $a$ and $b$ depend on the aggregate income and need, but not on their distributions. Let $i \in N$. By (\ref{eq1}),
$$
R_i \left( \left( \frac{Y}{n}, \ldots, \frac{Y}{n} \right), z \right) = \frac{Y}{n} + \left[z_i-\frac{Z}{n} \right] b(Y,Z).
$$
By \emph{core}, 
$$
R_i \left( \left( \frac{\rho Y}{n}, \ldots, \frac{\rho Y}{n} \right), \rho z \right) = \rho R_i \left( \left( \frac{Y}{n}, \ldots, \frac{Y}{n} \right), z \right),
$$
for each $\rho \in \mathbb{R}_{++}$. Equivalently,
$$
\left[z_i-\frac{Z}{n} \right] b(\rho Y,\rho Z) = \left[z_i-\frac{Z}{n} \right] b(Y,Z).
$$
That is, $b(Y,Z)$ is a homogeneous function of degree zero. Let us define $B:\mathbb{R}\to\mathbb{R}$ 
as $B\left( \frac{Y}{Z} \right) = b \left( \frac{Y}{Z},1 \right)$. Similarly, we define $A:\mathbb{R}\to\mathbb{R}$, as $A\left( \frac{Y}{Z} \right) = a \left( \frac{Y}{Z},1 \right)$. Therefore,
$$
R_i(y,z)= \frac{Y}{n} +  \left[y_i-\frac{Y}{n} \right] A\left( \frac{Y}{Z} \right) + \left[z_i-\frac{Z}{n} \right] B\left( \frac{Y}{Z} \right).
$$
Finally, note that $A$ and $B$ are continuous because so are $a$ and $b$. 
\qed
\subsection*{Proof of Theorem \ref{NAT+ST}}
We start showing that each rule defined as in the statement satisfies all the axioms therein. By Theorem \ref{thm_AB}, this happens with \textit{core} and \textit{no advantageous transfer}. As for \textit{stability}, $R^L$ trivially satisfies it, while for the other rules, for each $(y,z) \in \mathcal{D}^N$, and each $i \in N$,
$$
R_i \left( R_i(y,z),z \right) = \frac{\sum_{i=1}^n R_i(y,z)}{n} + \left[z_i-\frac{Z}{n} \right] B \left( \frac{\sum_{i=1}^n R_i(y,z)}{Z} \right) = \frac{Y}{n} + \left[z_i-\frac{Z}{n} \right] B \left( \frac{Y}{Z} \right) = R_i(y,z).
$$
Now, we prove the converse implication. Let $R$ be a rule that satisfies the axioms in the statement. By Theorem \ref{thm_AB}, there exist two functions $A,B: \mathbb{R} \longrightarrow \mathbb{R}$, such that, for each $(y,z) \in \mathcal{D}^N$ and each $i \in N$,
$$
R_i(y,z)= \frac{Y}{n} +  \left[y_i-\frac{Y}{n} \right] A \left( \frac{Y}{Z} \right) + \left[z_i-\frac{Z}{n} \right] B \left( \frac{Y}{Z} \right).
$$
Notice that, by definition, $\sum_{i \in N} y_i = Y = \sum_{i \in N} R_i(y,z)$. Then, using the previous expression,
\begin{align*}
R_i \left( R(y,z), z \right) &= \frac{Y}{n} +  \left[R_i(y,z)-\frac{Y}{n} \right] A \left( \frac{Y}{Z} \right) + \left[z_i-\frac{Z}{n} \right] B \left( \frac{Y}{Z} \right) \\
&= \frac{Y}{n} +  \left[\frac{Y}{n} +  \left[y_i-\frac{Y}{n} \right] A \left( \frac{Y}{Z} \right) + \left[z_i-\frac{Z}{n} \right] B \left( \frac{Y}{Z} \right) -\frac{Y}{n} \right] A \left( \frac{Y}{Z} \right) + \left[z_i-\frac{Z}{n} \right] B \left( \frac{Y}{Z} \right) \\
&= \frac{Y}{n} +  \left[y_i-\frac{Y}{n} \right] A \left( \frac{Y}{Z} \right)^2 + \left[z_i-\frac{Z}{n} \right] B \left( \frac{Y}{Z} \right) \left[ 1 + A \left( \frac{Y}{Z} \right) \right].
\end{align*}
\emph{Stability} requires that $R(R(y,z),z)) = R(y,z)$. Thus, 
$$
A \left( \frac{Y}{Z} \right)^2 = A \left( \frac{Y}{Z} \right),
\text{ and }
B \left( \frac{Y}{Z} \right) \left[ 1 + A \left( \frac{Y}{Z} \right) \right] = B \left( \frac{Y}{Z} \right).
$$
The solutions of the previous system of functional equations satisfy either $A \left( \frac{Y}{Z} \right) \equiv 0$ or  $A \left( \frac{Y}{Z} \right) \equiv 1$. When $A \left( \frac{Y}{Z} \right) \equiv 0$, $B \left( \frac{Y}{Z} \right)$ may take any value, while when $A \left( \frac{Y}{Z} \right) \equiv 1$, $B \left( \frac{Y}{Z} \right) \equiv 0$. Therefore, depending on the case, we obtain the expression of the two rules mentioned in the statement. \qed
\subsection*{Proof of Theorem \ref{NAT+D}}
We focus on the non-trivial implication. Suppose that $R$ satisfies all the axioms in the statement. By Theorem \ref{thm_AB}, there exist two functions $A,B: \mathbb{R} \longrightarrow \mathbb{R}$, such that $R \equiv R^{AB}$. Let $i \in N$ and $(y,z) \in \mathcal{D}^N$ be such that $y_i=z_i=0$. By \emph{dummy}, 
$$
0 = R_i(y,z) = \frac{Y}{n} -\frac{Y}{n} A \left( \frac{Y}{Z} \right) - \frac{Z}{n} B \left( \frac{Y}{Z} \right).
$$
Thus, 
$$
B \left( \frac{Y}{Z} \right) = \left[ 1- A \left( \frac{Y}{Z} \right) \right] \frac{Y}{Z}.
$$
Then, replacing function $B$ in the expression of $R^{AB}$, we have that, for each $j \in N$,
\begin{align*}
R_j(y,z) &= \frac{Y}{n} +  \left[y_j-\frac{Y}{n} \right] A \left( \frac{Y}{Z} \right) + \left[z_j-\frac{Z}{n} \right] \left[ 1- A \left( \frac{Y}{Z} \right) \right] \frac{Y}{Z} \\
&= A \left( \frac{Y}{Z} \right) y_j + \left[ 1- A \left( \frac{Y}{Z} \right)\right] \frac{z_j}{Z}Y.
\end{align*}
\qed
\subsection*{Proof of Proposition \ref{thm_LPF}}
We focus on the first statement.\footnote{As for the second statement, it would simply follow from the fact that the rules therein are dual of the rules in the first statement.} Theorem \ref{thm_AB} guarantees that the rules in such a statement satisfy \textit{core} and \textit{no advantageous transfer}. As for \textit{income additivity}, let $(y,z),(y',z') \in \mathcal{D}^{N}$, and let $i \in N$. Then,
\begin{align*}
R_i(y+y',z) &= \alpha_1 (y_i+y'_i) + \alpha_2 \frac{z_i}{Z} (Y+Y') + (1-\alpha_y-\alpha_z) \frac{Y+Y'}{n} \\
&= \alpha_1 y_i + \alpha_2 \frac{z_i}{Z} Y + (1-\alpha_1-\alpha_2) \frac{Y}{n}  + \alpha_1 y'_i + \alpha_2 \frac{z_i}{Z} Y' + (1-\alpha_1-\alpha_2) \frac{Y'}{n} \\
&= R_i(y,z) +R_i(y',z).
\end{align*}
Conversely, let $R$ be a rule that satisfies all the axioms in the (first) statement. By Theorem \ref{thm_AB}, there exist two functions $A,B: \mathbb{R} \longrightarrow \mathbb{R}$, such that, for each $(y,z) \in \mathcal{D}^N$, and each $i \in N$,
$$
R_i(y,z)= \frac{Y}{n} +  \left[y_i-\frac{Y}{n} \right] A \left( \frac{Y}{Z} \right) + \left[z_i-\frac{Z}{n} \right] B \left( \frac{Y}{Z} \right).
$$
On the one hand, \emph{income additivity} implies that
$$
R_i \left( \left( \frac{Y}{n}, \ldots, \frac{Y}{n} \right), z \right) + R_i \left( \left( \frac{Y'}{n}, \ldots, \frac{Y'}{n} \right), z \right) = R_i \left( \left( \frac{Y+Y'}{n}, \ldots, \frac{Y+Y'}{n} \right), z \right).
$$
And then,
$$
B \left( \frac{Y}{Z} \right) + B \left( \frac{Y'}{Z} \right) = B \left( \frac{Y+Y'}{Z} \right),
$$
which is, simply, Cauchy's equation. As $B$ is continuous, it follows that 
$B(t)=\alpha_2 t$ for some constant $\alpha_2 \in \mathbb{R}$ \citep[e.g.,][]{aczel1966lectures}. 
Therefore,
$$
R_i(y,z)= \frac{Y}{n} +  \left[y_i-\frac{Y}{n} \right] A \left( \frac{Y}{Z} \right) + \alpha_2 \frac{Y}{Z} \left[z_i-\frac{Z}{n} \right].
$$
On the other hand, let $(y,z),(y',z) \in \mathcal{D}^N$ be two problems with the same needs. Recall that the function $A$ depends on the aggregate income and need, but not on their distributions. Let $s \in \mathbb{R}^n$ be such that $s_j=y'_j-\frac{Y}{n}$ for all $j \in N$. Then, using the previous expression, we obtain that, for each $i\in N$,
$$
R_i \left( \left( \frac{Y}{n}, \ldots, \frac{Y}{n} \right), \left( \frac{Z}{n}, \ldots, \frac{Z}{n} \right) \right) = \frac{Y}{n},
$$
$$
R_i \left( y', \left( \frac{Z}{n}, \ldots, \frac{Z}{n} \right) \right) = R_i \left( \left( \frac{Y}{n}+s_1, \ldots, \frac{Y}{n}+s_n \right), \left( \frac{Z}{n}, \ldots, \frac{Z}{n} \right) \right) = \frac{Y'}{n} + \left[ \frac{Y}{n} + s_i - \frac{Y'}{n} \right] A \left(\frac{Y'}{Z} \right),
$$
and 
\begin{align*}
R_i \left( \left( \frac{Y}{n}, \ldots, \frac{Y}{n}\right)+y', \left( \frac{Z}{n}, \ldots, \frac{Z}{n} \right) \right) &= R_i \left( \left( \frac{Y}{n} + \frac{Y}{n}+s_1, \ldots, \frac{Y}{n} + \frac{Y}{n}+s_n \right), \left( \frac{Z}{n}, \ldots, \frac{Z}{n} \right) \right) \\
&= \frac{Y+Y'}{n} + \left[ \frac{Y}{n} + \frac{Y}{n} + s_i - \frac{Y+Y'}{n} \right] A \left(\frac{Y+Y'}{Z} \right) \\
&= \frac{Y+Y'}{n} + \left[\frac{Y}{n} + s_i - \frac{Y'}{n} \right] A \left(\frac{Y+Y'}{Z} \right).
\end{align*}
By \emph{income additivity}, 
$$
R_i \left( \left( \frac{Y}{n}, \ldots, \frac{Y}{n}\right)+y', \left( \frac{Z}{n}, \ldots, \frac{Z}{n} \right) \right) = R_i \left( \left( \frac{Y}{n}, \ldots, \frac{Y}{n} \right), \left( \frac{Z}{n}, \ldots, \frac{Z}{n} \right) \right) + R_i \left( y', \left( \frac{Z}{n}, \ldots, \frac{Z}{n} \right) \right).
$$
Or, equivalently,
$$
A \left(\frac{Y'}{Z} \right) = A \left(\frac{Y'}{Z} + \frac{Y}{Z} \right).
$$
Thus, $A$ is a continuous function that satisfies the condition $A(t)=A(t+u)$ for each pair $t,u \in \mathbb{R}$. Thus, $A$ is a constant function, i.e., $A(t)=\alpha_1 \in \mathbb{R}$. Therefore,
\begin{align*}
R_i(y,z) &= \frac{Y}{n} +  \alpha_1 \left[y_i-\frac{Y}{n} \right] + \alpha_2 \frac{Y}{Z} \left[z_i-\frac{Z}{n} \right] \\
&= \alpha_1 y_i + \alpha_2 \frac{z_i}{Z} Y + (1-\alpha_1-\alpha_2) \frac{Y}{n},
\end{align*}
as desired.
\qed

\subsection*{Proof of Proposition \ref{thm_PF}}
By the first statement of Proposition \ref{thm_LPF}, a rule $R$ satisfies \emph{core}, \emph{no advantageous transfer}, \emph{stability} and \emph{income additivity} if and only if, for each  
$(y,z) \in \mathcal{D}^N$ and each $i \in N$, 
$$
R_i(y,z)=\alpha_1 y_i + \alpha_2 \frac{z_i}{Z} Y + (1-\alpha_1-\alpha_2) \frac{Y}{n}.
$$
In particular, as $Y=\sum_{i \in N} R_(y,z)$, 
$$R_i(R_i(y,z),z)=\alpha_1 R_i(y,z) + \alpha_2 \frac{z_i}{Z} Y + (1-\alpha_1-\alpha_2) \frac{Y}{n}.$$
Altogether, we have that $R$ satisfies \emph{stability}, i.e., $R_i(R_i(y,z),z)=R_i(y,z)$, if and only if 
either $R_i(y,z)=y_i$, or $\alpha_1=0$. In the former case, $R$ is \emph{laissez-faire}. In the latter case, $R$ is a convex combination between \emph{full redistribution} and \emph{proportional redistribution}, as desired. The second statement follows from duality.\qed


\subsection*{Proof of Proposition \ref{LP-cvx}}
This is a direct consequence of the fact that the rules in the statements of Proposition \ref{thm_LPF} satisfy \textit{dummy} if and only if $\alpha_1+\alpha_2=1$. 


\end{document}